\documentclass[aps,twocolumn,superscriptaddress]{revtex4}

\usepackage{amsmath,amssymb,graphicx}
\usepackage{color}

\graphicspath{{fig-ps/}}

\begin{document}

\title{
Dark spherical shell solitons in three-dimensional Bose-Einstein condensates:
Existence, stability and dynamics}

\author{Wenlong Wang}
\email{wenlong@physics.umass.edu}
\affiliation{Department of Physics and Astronomy, Texas A$\&$M University,
College Station, Texas 77843-4242, USA}
\affiliation{Department of Physics, University of Massachusetts,
Amherst, Massachusetts 01003 USA}

\author{P.G. Kevrekidis}
\email{kevrekid@math.umass.edu}
\affiliation{Department of Mathematics and Statistics, University of Massachusetts,
Amherst, Massachusetts 01003-4515 USA}
\affiliation{Center for Nonlinear Studies and Theoretical Division, Los Alamos
National Laboratory, Los Alamos, NM 87544}

\author{R. Carretero-Gonz{\'a}lez}
\affiliation{Nonlinear Dynamical Systems
Group,\footnote{\texttt{URL}: http://nlds.sdsu.edu}
Computational Sciences Research Center, and
Department of Mathematics and Statistics,
San Diego State University, San Diego, California 92182-7720, USA}

\author{D. J. Frantzeskakis}
\affiliation{Department of Physics, University of Athens,
Panepistimiopolis, Zografos, Athens 15784, Greece}

\begin{abstract}
In this work we study
spherical shell dark soliton states in three-dimensional
atomic Bose-Einstein condensates.
Their symmetry is exploited in order to analyze
their existence,
as well as that of topologically charged variants of the structures, and,
importantly, to identify their
linear stability Bogolyubov-de Gennes spectrum.
We compare our effective 1D spherical and 2D cylindrical computations
with the full 3D numerics. An important conclusion is that such
spherical shell solitons can be stable sufficiently close to the linear limit
of the isotropic condensates considered herein. We have also
identified their instabilities leading to
the emergence of vortex line and vortex ring cages.
In addition, we generalize effective particle pictures of lower dimensional
dark solitons and ring dark solitons
to the spherical
shell solitons concerning
their equilibrium radius and effective dynamics around it. In this case too, we
favorably compare the resulting predictions
such as the shell equilibrium radius,
qualitatively and quantitatively, with full
numerical solutions in 3D.
\end{abstract}

\maketitle

\section{Introduction}

The pristine 
setting of atomic Bose-Einstein condensates (BECs)
has offered a significant playground for the examination of numerous
physical concepts~\cite{book1,book2}. One of the flourishing directions has been
at the interface of the theory of nonlinear waves and such
atomic (as well as optical) systems, concerning, in particular, the study of matter-wave
solitons~\cite{emergent,book_new}. Such coherent nonlinear structures have been observed in experiments
and studied extensively in theory.
Prototypical examples of pertinent studies include ---but are not limited to---
bright~\cite{expb1,expb2,expb3},
dark~\cite{djf} and gap~\cite{gap} matter-wave solitons, as well as
vortices~\cite{fetter1,fetter2}, 
solitonic vortices and vortex rings~\cite{komineas_rev}.

Among these diverse excitations, dark solitons in repulsive BECs
have enjoyed
a considerable amount of attention in effectively one-dimensional
settings due to numerous experiments leading
(especially, more recently) to their well-controlled
creation~\cite{chap01:denschl,chap01:dark1,ourmarkus2,ourmarkus3,seng1,seng2}.
Moreover, numerous works considered dark solitons in
higher-dimensional settings in order to experimentally
explore their instability leading to the formation of
vortex rings and vortex lines, as illustrated, e.g.,
in Refs.~\cite{chap01:dark,engels,jeff,rcg:90,pitasv}. More recently,
such excitations have also been observed
in fermionic
superfluids~\cite{martinz,fprl}.

On the other hand, multi-dimensional (i.e., non-planar)
variants of 
dark solitons, were studied as well ---cf.~e.g., Chapters 7 and 8 of Ref.~\cite{emergent}.
Examples of such structures are radially symmetric solitons, such as
the ring dark solitons (RDS),
initially proposed in atomic BECs in Ref.~\cite{ring_pap}
(see also Refs.~\cite{carrc,GregHerring,middel10a} for subsequent studies).
Nevertheless, such structures were studied to a far lesser extend,
arguably 
due to their generic identified instability, which
was found to give rise to states such as alternating
charge vortex polygons.
Notice that, as was recently shown in Ref.~\cite{wenlong}, 
stabilization of RDS
is possible upon employing 
a proper 
potential barrier.

The three-dimensional non-planar generalization of
dark solitons,
namely spherical shell solitons, has
received very limited attention. The analysis of Ref.~\cite{carrc}
in the case of box potentials (rather than the typical parabolic
trap setting of BECs) revealed their potential existence, but
did not pursue their stability analysis. Moreover, the adiabatic
dynamics of these structures was studied analytically via a variational
approach
in Ref.~\cite{gl} without providing, however, relevant numerical results.
Nevertheless, it is important to note that in an
experimental realization of vortex ring states, 
the transient observation of spherical shell solitons was reported~\cite{harvard}. 
In addition, they were also suggested as a potential outcome of collisions
of vortex rings in the work of Ref.~\cite{kominbrand}.

It is 
the purpose of the present manuscript to revisit
these states and explore not only their existence in isotropic traps,
but also perform 
their systematic Bogolyubov-de Gennes (BdG) stability analysis.
This is done by projecting perturbations onto eigenfunctions
of the linear problem in the angular directions.
Using this basis, 
the relevant computations are greatly simplified, and
the full 3D spectrum computation is reduced to an
effectively small set of one-dimensional problems (for the
different modes). It should be mentioned
here that the use of spherical harmonics as a basis in the
angular variables for states bearing radial symmetry is
rather common in terms of numerical schemes for linear and nonlinear
Schr{\"o}dinger (NLS) equations ---see, e.g., Ref.~\cite{hayder} for a
a recent example.
Earlier efforts along these lines in the context of linear
Schr{\"o}dinger equations can be found, e.g., in the
works of Refs.~\cite{herman,corey}. In 2D BECs, analogous decompositions
of azimuthal modes for radial states including RDSs
and vortices~\cite{carrc,GregHerring,kollar}
have appeared, and in 3D similar possibilities are starting
to emerge~\cite{wuming}.

In 
this work, we find that indeed spherical shell solitons can be
identified as stable 3D solutions of the Gross-Pitaevskii
equation, sufficiently
close to the linear (small-amplitude) limit of small chemical potential
from which they bifurcate, as is shown below. Moreover, we provide
a theoretical analysis of their dynamics as ``effective particles''.
This approach allows to characterize the opposite limit of
large chemical potential. Our numerical computations of existence
and stability, complemented by direct numerical simulations of
the spherical shell soliton dynamics, not only corroborate the above
two limits, but also provide a systematic way to interpolate
between the two analytically tractable regimes.

Our presentation 
is organized as follows. 
First, in Sec.~\ref{setup}, we introduce
the model and describe the implementation of the linear stability analysis.
Our analytical approaches for the spherical shell dark solitons
in the above mentioned analytically tractable limits
are then presented in Sec.~\ref{sec:sub:part}. Next, in Sec.~\ref{results},
we present our systematic numerical results.
Finally, our conclusions and a number of open
problems for future consideration are given in Sec.~\ref{conclusion}.
In the Appendix, a similar numerical decomposition method
but using cylindrical (rather than spherical) symmetry is also
discussed whose results are also given for comparison
in Sec.~\ref{results}.

\section{Model and Computational Setup}
\label{setup}

\subsection{The Gross-Pitaevskii equation}

In the framework of lowest-order mean-field theory, and for sufficiently
low-temperatures, the dynamics of a 3D repulsive BEC,
confined in a time-independent trap $V$, is described by the following
dimensionless Gross-Pitaevskii equation (GPE)~\cite{book1,book2,emergent,book_new}:
\begin{equation}
i \psi_t=-\frac{1}{2} \nabla^2 \psi+V \psi +| \psi |^2 \psi-\mu \psi,
\label{GPE}
\end{equation}
where $\psi(x,y,z,t)$ is the macroscopic wavefunction of the BEC and
$\mu$ is the chemical potential (subscripts denote partial derivatives).
Here, we consider a harmonic trap of the form:
\begin{equation}
V=\frac{1}{2} \omega_{\rho}^2 \rho^2+\frac{1}{2} \omega_z^2 z^2,
\label{potential}
\end{equation}
where $\rho=\sqrt{x^2+y^2}$, $\omega_{\rho}$ and $\omega_z$ are the
trapping frequencies along the $(x,y)$ plane and 
the vertical direction $z$, respectively.
Note that the potential has rotational symmetry with respect to the $z$-axis.
In our numerical simulations, we focus on the fully symmetric (isotropic,
spherically symmetric) case with
$\omega=\omega_{\rho}=\omega_z=1$, in which we can benchmark our numerical
methods in effective 1D and 2D against those of the fully 3D setting.

We examine the dark spherical shell solitons, hereafter
referred to as DSS. This single radial node state exists in this
isotropic case from the linear limit onwards as a stationary state of the form
$\psi(x,y,z,t)=e^{-i \mu t} \psi_{\rm{DSS}}(r)$
where $r=\sqrt{x^2+y^2+z^2}$ is the spherical radial variable.
At the linear limit, the waveform is an eigenmode of
the quantum harmonic oscillator with chemical potential $\mu=7\omega/2$,
and spatial profile:
\begin{eqnarray}
|\psi_{\rm{DSS}}\rangle_{\rm{linear}}&=&
\frac{1}{\sqrt{3}}\left(|200\rangle+|020\rangle+|002\rangle \right) \\
&\propto& \left(\omega r^2-\frac{3}{2}\right) e^{-\omega r^2/2},
\end{eqnarray}
where the basis $\{|n_x n_y n_z \rangle \}$ denotes
the 3D harmonic oscillator quantum states
in Cartesian coordinates.

The DSS is generally expected to be unstable for high
chemical potentials due to transverse modulational (snaking-type) instability,
in a way similar to its planar and RDS 
counterparts,
as summarized, e.g., in Refs.~\cite{book_new,djf}.
However, for smaller chemical potentials, and especially near
the linear limit, it is relevant to
analyze stability of the DSS 
and identify corresponding instabilities and where they may arise.
In what follows 
(cf.~Sec.~\ref{results} below),
we will turn to numerical computations,
based on a fixed point iteration and our spherical harmonic decomposition
of the spectral stability BdG problem,
in order to efficiently identify the intervals of existence and stability of
the relevant modes.

\subsection{The Linear Stability Problem: Spherical Harmonic Decomposition}

In this section, we discuss our analysis of
the linear stability of the DSS states
utilizing their effective 1D nature, namely their spherical symmetry.
In spherical coordinates $(r,\theta,\phi)$, the Laplacian can be decomposed into
radial and angular parts, denoted by $\Delta_R$ and $\Delta_S$ respectively, namely
\begin{equation}
\nabla^2 f =\Delta_R f+\frac{\Delta_S f}{r^2},
\end{equation}
where
%
\begin{eqnarray}
\Delta_R f &=&\frac{1}{r^2} \frac{\partial}{\partial r} \left(r^2 \frac{\partial f}{\partial r}\right),\\
\Delta_S f &=&\frac{1}{\sin\theta} \frac{\partial}{\partial \theta} \left(\sin\theta \frac{\partial f}{\partial \theta}\right)+\frac{1}{\sin^2\theta}\frac{\partial^2 f}{\partial \phi^2},
\end{eqnarray}
%
%

The $\Delta_S$ operator has eigenstates given by the spherical harmonics
$\{Y_{\ell m}\}$ with eigenvalues $-\ell(\ell+1)$, i.e.,
\begin{equation}
\Delta_S Y_{\ell m} =-\ell(\ell+1)Y_{\ell m}.
\end{equation}
For states with spherical symmetry ($\ell=0$), the $\Delta_S$ part of the Laplacian is not
relevant for identifying the stationary state. Thus, the stationary
DSS state $\psi_0(r)$ of Eq.~(\ref{GPE}),
defined in the domain $r \in [0,\infty)$, satisfies the following radial
equation:
\begin{equation}
-\frac{1}{2} \Delta_R \psi_0+V(r)\psi_0 +| \psi_0 |^2 \psi_0-\mu \psi_0=0.
\end{equation}
%
%

Now, let us consider the linear stability (BdG) spectrum of such a stationary state.
We consider small-amplitude perturbations
expanded using the complete basis of $\{Y_{\ell m}\}$ (the spherical
harmonic eigen-basis)
as follows:
\begin{equation}
\psi(r,t)=\psi_0+\sum\limits_{\ell m} \left[ a_{\ell m}(r,t)Y_{\ell m}+b_{\ell m}^*(r,t)Y_{\ell m}^* \right],
\label{perturbation}
\end{equation}
where the space- and time-dependent coefficients $a_{\ell m}(r,t)$ and $b_{\ell m}(r,t)$
determine the radial and temporal evolution of the perturbation.
%
%
Then, substituting Eq.~(\ref{perturbation}) into 
Eq.~(\ref{GPE})
and retaining up to linear terms in the expansion,
one notes that all the $\ell$ modes are
mutually independent due to the spherical symmetry of the steady state.
For mode $\ell$, the coefficients $a$ and $b$ of the expansion obey the following evolution equations:
$$
\begin{array}{rcl}
i a_t&=&-\frac{1}{2} \Delta_R a+ \frac{\ell(\ell+1)}{2r^2}a +Va + 2 |\psi_0|^2 a+
\psi_0^2 b-\mu a,\\[1.0ex]
-i b_t&=&-\frac{1}{2} \Delta_R b+ \frac{\ell(\ell+1)}{2r^2}b +Vb +2 |\psi_0|^2b
+ \psi_0^{*2}a-\mu b,
\end{array}
$$
where we have dropped for simplicity the subscripts $(\ell,m)$. 
As we are interested in the stability of the steady-state, we now expand
the coefficients using
\begin{equation}
\begin{pmatrix}
a\\
b
\end{pmatrix}
=
\begin{pmatrix}
a_0\\
b_0
\end{pmatrix}
e^{\lambda t},
\end{equation}
where $\lambda$ is an eigenvalue of the following matrix
\begin{equation}
M=
\begin{pmatrix}
M_{11} & M_{12} \\
M_{21} & M_{22}
\end{pmatrix},
\label{M}
\end{equation}
with
$$
\begin{array}{rcl}
M_{11}&=&-i\left(-\frac{1}{2} \Delta_R + \frac{\ell(\ell+1)}{2r^2}  +V  +2|\psi_0|^2-\mu\right),\\[2.0ex]
M_{12}&=&-i\psi_0^2,\\[1.0ex]
M_{21}&=&i\psi_0^{*2},\\[1.0ex]
M_{22}&=&i\left(-\frac{1}{2} \Delta_R + \frac{\ell(\ell+1)}{2r^2}  +V  +2|\psi_0|^2-\mu\right).
\end{array}
$$
One can, therefore, compute the full spectrum by computing that of
each mode $\ell$
independently, and then putting them all together to ``reconstruct'' the
full spectrum. In what follows we use
$\ell=0,1,2,...,10$ for values of the chemical potential up to $\mu \approx 12$.
The reduction of dimensionality with the decomposition on a complete basis set
is also available in 2D for states with rotational symmetry up to a topological
charge along the $z$ axis. For clarity, this is shown in the Appendix. It is worth mentioning
that a large class of BEC states belongs to this class.

Finally, we discuss
the application of the degenerate perturbation method (DPM)~\cite{ourVRfromLinear}
to the DSS state. This method is employed for the determination of the DSS' spectrum
%
near the linear limit, 
and also for the identification of the nature of 
instabilities.
The method can be most readily
understood from the 3D version of the matrix $M$, where one can separate the
free (linear) part with known eigenstates and eigenvalues and the rest of the terms
can be treated as perturbations near the linear limit. Note that there are
2$\times$2 block matrices within $M$
and the complete basis consists of both ``positive''
and ``negative'' eigenstates. This terminology is associated with
quantum harmonic oscillator eigenstates
leading to positive or negative eigenvalues for the linear analogs of
the operators $M_{ii}$ in Eq.~(\ref{M}) i.e., with $|\psi_0|^2$ set to $0$.
In DPM, the functional dependence of the eigenvalues on
the chemical potential $\mu$ is dominated by the
interplay of equi-energetic, degenerate states. The
result of the nonlinear perturbations ($\propto |\psi_0|^2$)
in $M_{ij}$ within
Eq.~(\ref{M}) is to cause the
degenerate eigenvalues to depart from their respective linear limit, as
the norm of the solution (mass) increases. The free part provides the
basis that is formed from ``up'' states and ``down'' states, of the form
$
\begin{pmatrix}
|n_xn_yn_z \rangle \\
0
\end{pmatrix}
$
and
$
\begin{pmatrix}
0 \\
|n_xn_yn_z \rangle
\end{pmatrix},
$
with eigenvalues $E_{|n_xn_yn_z \rangle}-E_{\rm{DSS}}$ and $E_{\rm{DSS}}-E_{|n_xn_yn_z \rangle}$ respectively, where $E_\Lambda$ is the eigenenergy of the 3D 
quantum harmonic oscillator for eigenstate $\Lambda$. The relevant states for eigenvalues
near the spectrum of Im($\lambda$)=0, 1 and 2 are listed in Table.~\ref{para}.
Then, the effect of the nonlinearity on these eigenmodes and the
corresponding eigenvalue corrections are calculated, as discussed
in Ref.~\cite{ourVRfromLinear} (and originally spearheaded in
this context in Ref.~\cite{feder}). Below, this is compared to
the numerical spectrum in Sec.~\ref{results}.

\begin{table}
\caption{
Relevant states for eigenvalues near the spectrum of Im($\lambda$)=0,1 and 2 for the degenerate perturbation theory. Here $|m\,n\,p \rangle$ stands for eigenstates with all possible distinct permutations with the quantum numbers $m,n$ and $p$.
\label{para}
}
\begin{tabular}{ccc}
\hline
\hline
Im($\lambda$)  & ``up'' states  & ``down'' states \\
\hline
$0$  & $|002 \rangle$ $|011 \rangle$ & $|002 \rangle$ $|011 \rangle$ \\
\hline
$1$  & $|003 \rangle$ $|012 \rangle$ $|111 \rangle$& $|001 \rangle$ \\
\hline
$2$  &\qquad $|004 \rangle$ $|013 \rangle$ $|022 \rangle$ $|112 \rangle$ \qquad & $|000 \rangle$ \\
\hline
\hline
\end{tabular}
\end{table}

\section{The particle picture for the DSS}
\label{sec:sub:part}

We now turn to the limit of large chemical potential $\mu$.
In this so-called
Thomas-Fermi (TF) limit of large $\mu$,
the ground state of the GPE is approximated as 
$\psi_{\rm{TF}}=\sqrt{\max(\mu-V,0)}$ \cite{book1,book2}. A natural way to obtain
a reduced dynamical description of the DSS
---analogously to what is done in lower-dimensional settings---
is to develop an effective particle picture
for the DSS in the TF limit. Here, following
Ref.~\cite{wenlong}, 
we focus on the equilibrium radius $r_c$ of the DSS, as a prototypical diagnostic that we can compare
to our numerical results. This equilibrium radius is determined as the critical value
of the DSS radius for which the restoring force due to the harmonic trap
is counterbalanced by the effective force
exerted due to the curvature of the DSS.

\begin{figure}
\begin{center}
\includegraphics[width=8cm]{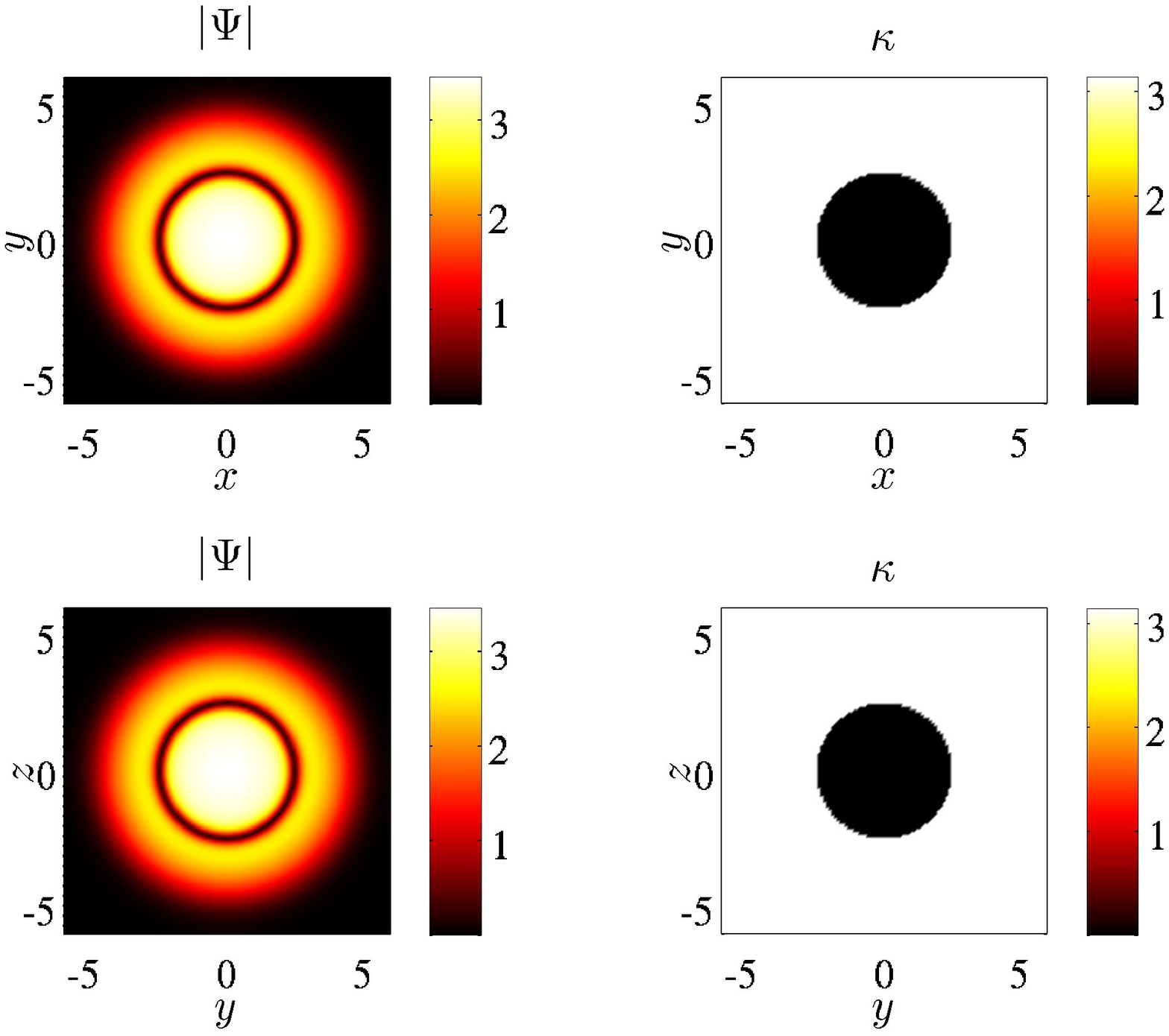}
\includegraphics[width=6cm]{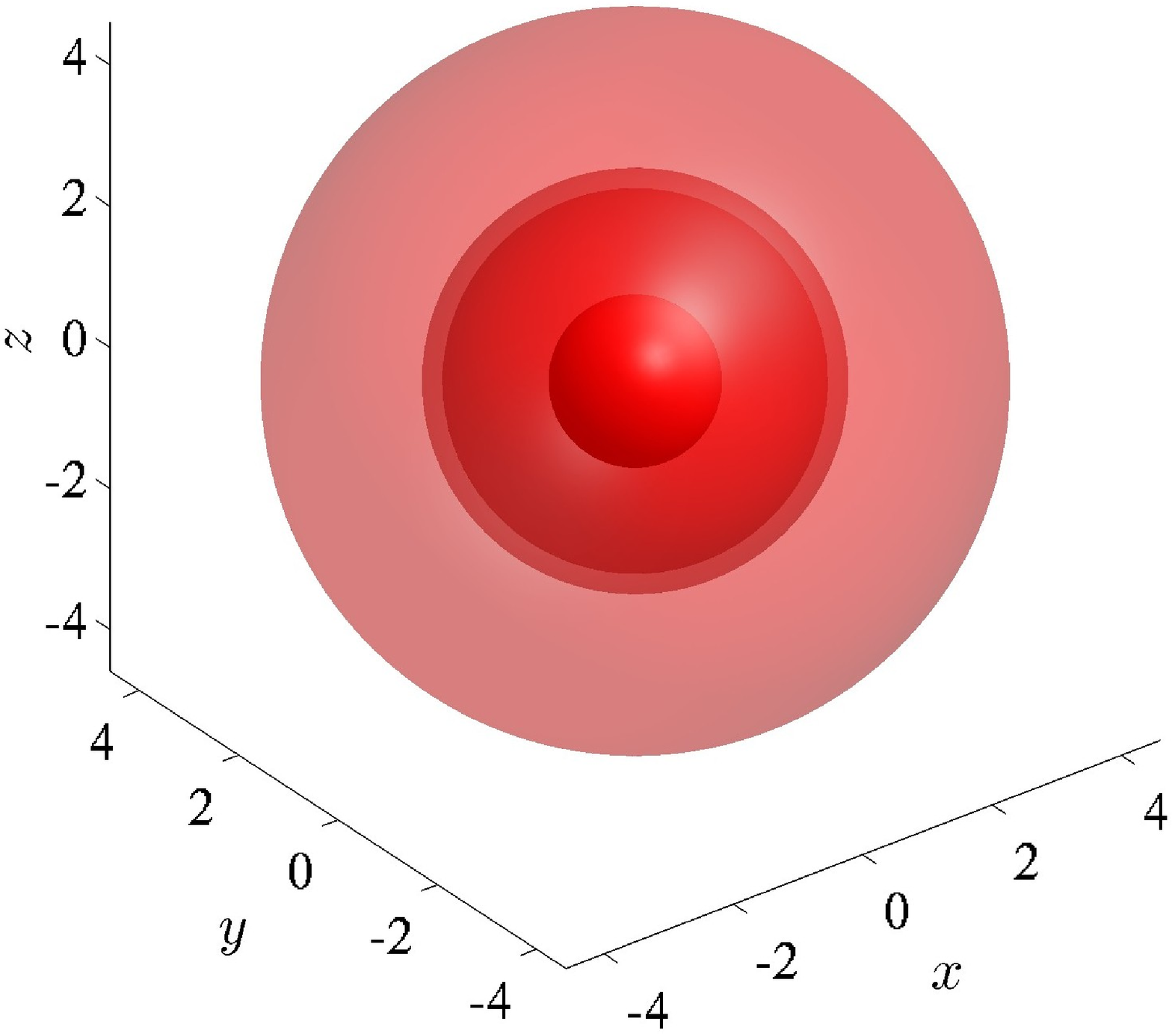}
\caption{(color online) The two pairs of the top panels show the
modulus $|\Psi|$ (left panels) and argument $\kappa$ (right panels)
of a dark soliton shell state at $\mu=12$; the top row illustrates the $z=0$ plane, while
the middle row the $x=0$ plane.
The bottom panel shows an isocontour plot of the same state.
}
\label{state}
\end{center}
\end{figure}

\begin{figure}[tb]
\begin{center}
\includegraphics[width=8cm]{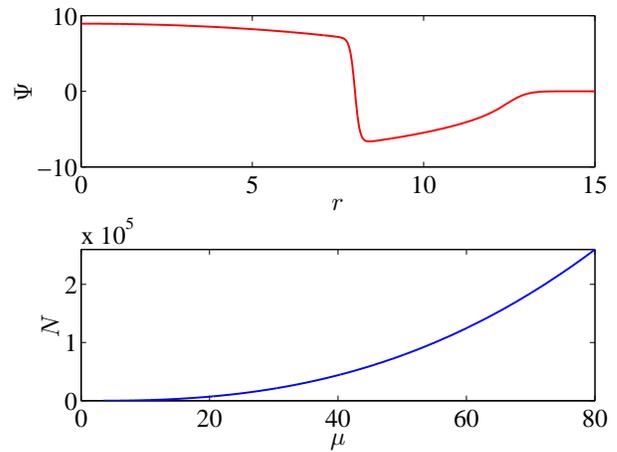}
\caption{
(Color online) Top panel: a radial profile of the dark soliton shell state in the TF limit for $\mu=80$.
Bottom panel: number of particles
as a function of chemical potential $\mu$, showing the
continuation of states from the linear limit to the nonlinear regime.
}
\label{LC}
\end{center}
\end{figure}

The first approach (motivated also by the work of Ref.~\cite{kapermorg}) focuses on
the stationary state using the ansatz
$\psi(r)=\psi_{\rm TF}(r) q(r)$, where one obtains an equation
for the DSS profile $q(r)$ 
(on top of the TF ground-state $\psi_{\rm TF}$) given by:
\begin{eqnarray}
\frac{1}{2} q'' + \mu q (1-q^2) = P(r),
\label{neweq}
\end{eqnarray}
where
$$
P(r)= V q (1-q^2) -\frac{2}{r} q'
-\frac{\psi_{\rm TF}''}{2 \psi_{\rm TF}} q
- \frac{\psi_{\rm TF}'}{\psi_{\rm TF}} q'
- \frac{1}{r} \frac{\psi_{\rm TF}'}{\psi_{\rm TF}} q,
$$
and primes denote derivatives with respect to $r$. Inspired by the
profile of the 1D and 2D equivalents of the DSS, namely the dark soliton and the RDS,
we seek a stationary DSS solution in the form of
$q(r)=\tanh[\sqrt{\mu} (r-r_c)]$.
Multiplying both sides of Eq.~(\ref{neweq}) by $q'$ and integrating over $r$
from $-\infty$ to $\infty$ (bearing in mind that the contribution of the
integral from $r=-\infty$ to $r=0$ is exponentially small), we find that
the equilibrium radius is given by
\begin{eqnarray}
r_c=\frac{\sqrt{\alpha\mu}}{\omega},
\label{neweq3}
\end{eqnarray}
where $\alpha=5-\sqrt{17} \approx 0.8769$. This result, as in the case of
the dark soliton ring in 2D~\cite{wenlong}, slightly overestimates
the actual value for $\alpha$ that, as we will see below, is accurately
predicted by the second particle picture approach, which retrieves the
precise value for $\alpha$ to be $\alpha=4/5$; this is confirmed by
the numerical results shown below, in Sec.~\ref{results}.

The second approach relies on energy conservation and it is based on the analysis
of Ref.~\cite{kamch}. In this approach, it is argued that the equation of
motion can be derived by a local conservation law (i.e., an adiabatic
invariant) in the form of the energy of a dark soliton under the
effect of curvature and of the density variation associated with it.
More specifically, knowing that the energy of the 1D dark
soliton centered at $x_c$
is given by ${\cal E}=(4/3) (\mu - \dot{x}_c )^{3/2}$~\cite{djf},
the generalization of the relevant adiabatic invariant quantity in a 3D
domain bearing density modulations reads:
\begin{equation}
\begin{array}{rcl}
{\cal E}&=&4 \pi r^2 \left[ \frac{4}{3} (\mu-V(r)-\dot{r}^2)^{3/2} \right]
\\[1.0ex]
        &=&4 \pi r_0^2 \left[ \frac{4}{3} (\mu-V(r_0)-\dot{r}_0^2)^{3/2}
\right],
\end{array}
\label{PPnew1}
\end{equation}
where $r(t)$ and $\dot{r}(t)$ are the (radial) location and velocity of the
DSS with initial conditions $r_0$ and $\dot{r}_0$.
Taking a time derivative on both sides of Eq.~(\ref{PPnew1}) and assuming
that the DSS has no initial speed ($\dot{r}_0=0$), we obtain a
Newtonian particle equation of motion
for the DSS of the following form:
\begin{equation}
\ddot{r}=-\frac{1}{2}\frac{\partial V}{\partial r}+\frac{2}{3 r}\left(\frac{r_0}{r}\right)^{4/3}[\mu-V(r_0)].
\label{PPextra}
\end{equation}
From this equation, the equilibrium position for the DSS yields
$r_c=\sqrt{\dfrac{4\mu}{5}}/\omega$, a result 
consistent with the numerical findings
of the next Section.
We now proceed to test these predictions, the stability analysis spectrum,
and the equilibrium positions for the DSS.

\begin{figure}[t]
\begin{center}
\includegraphics[width=7cm]{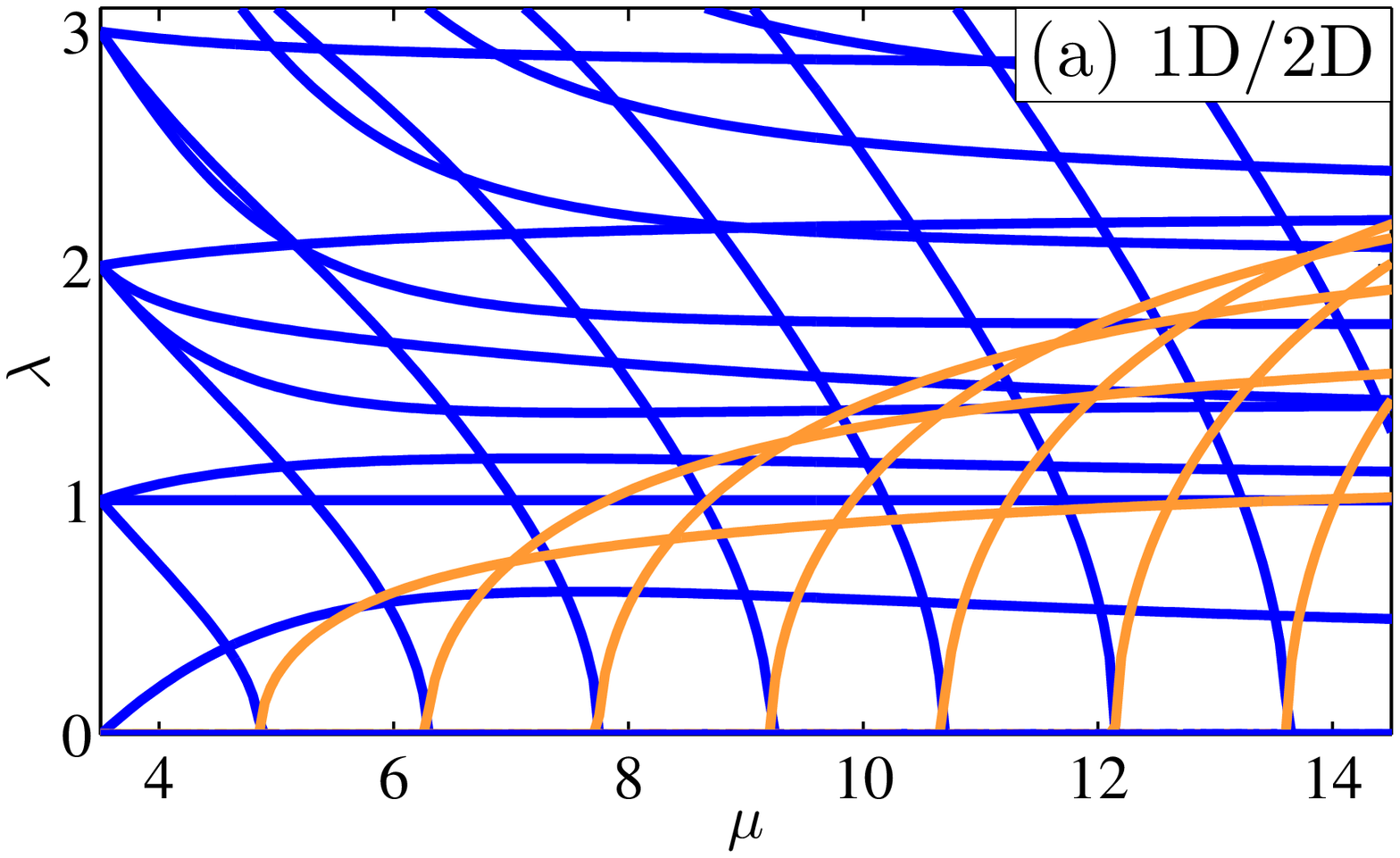}
\includegraphics[width=7cm]{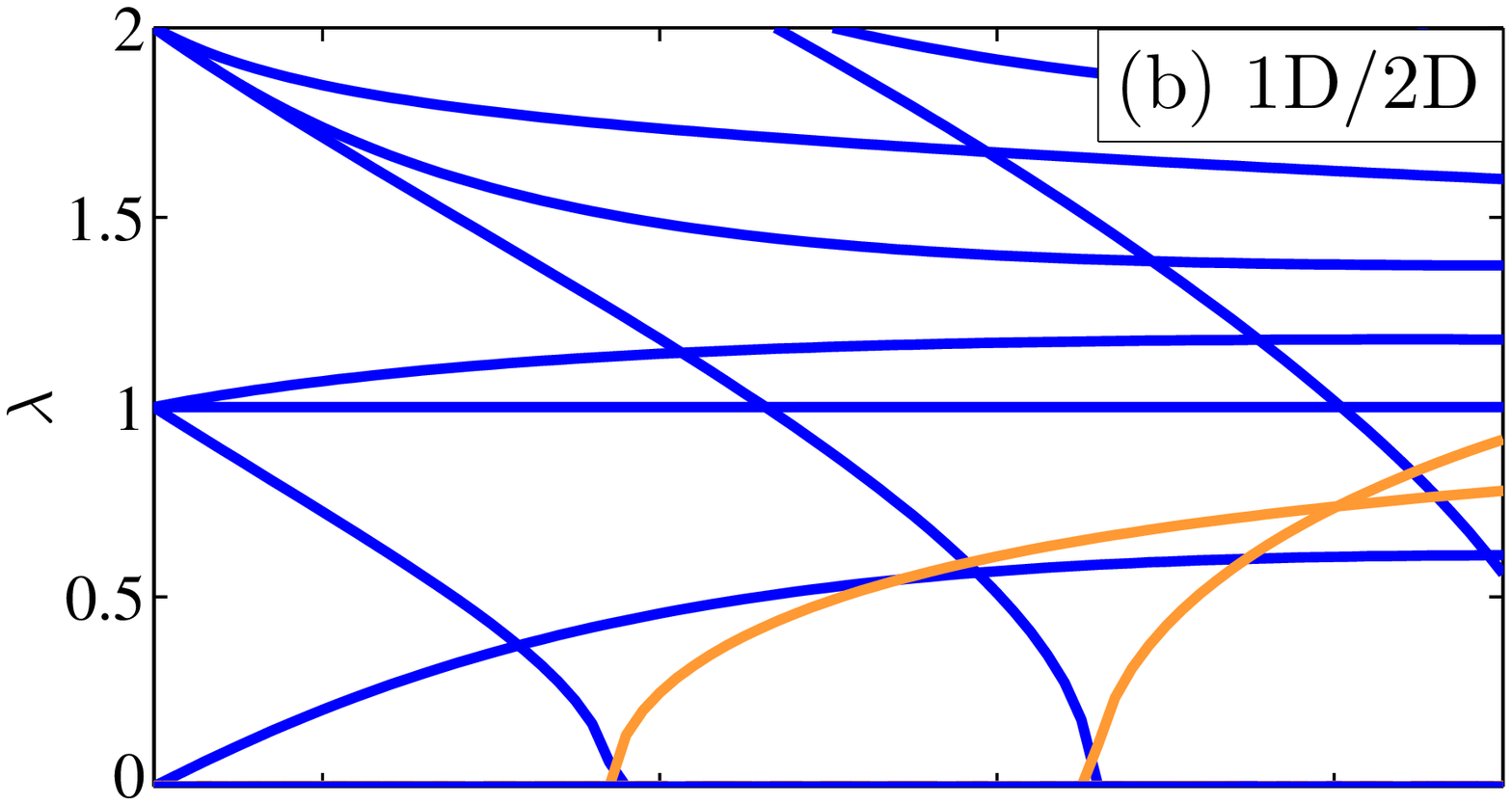}
\includegraphics[width=7cm]{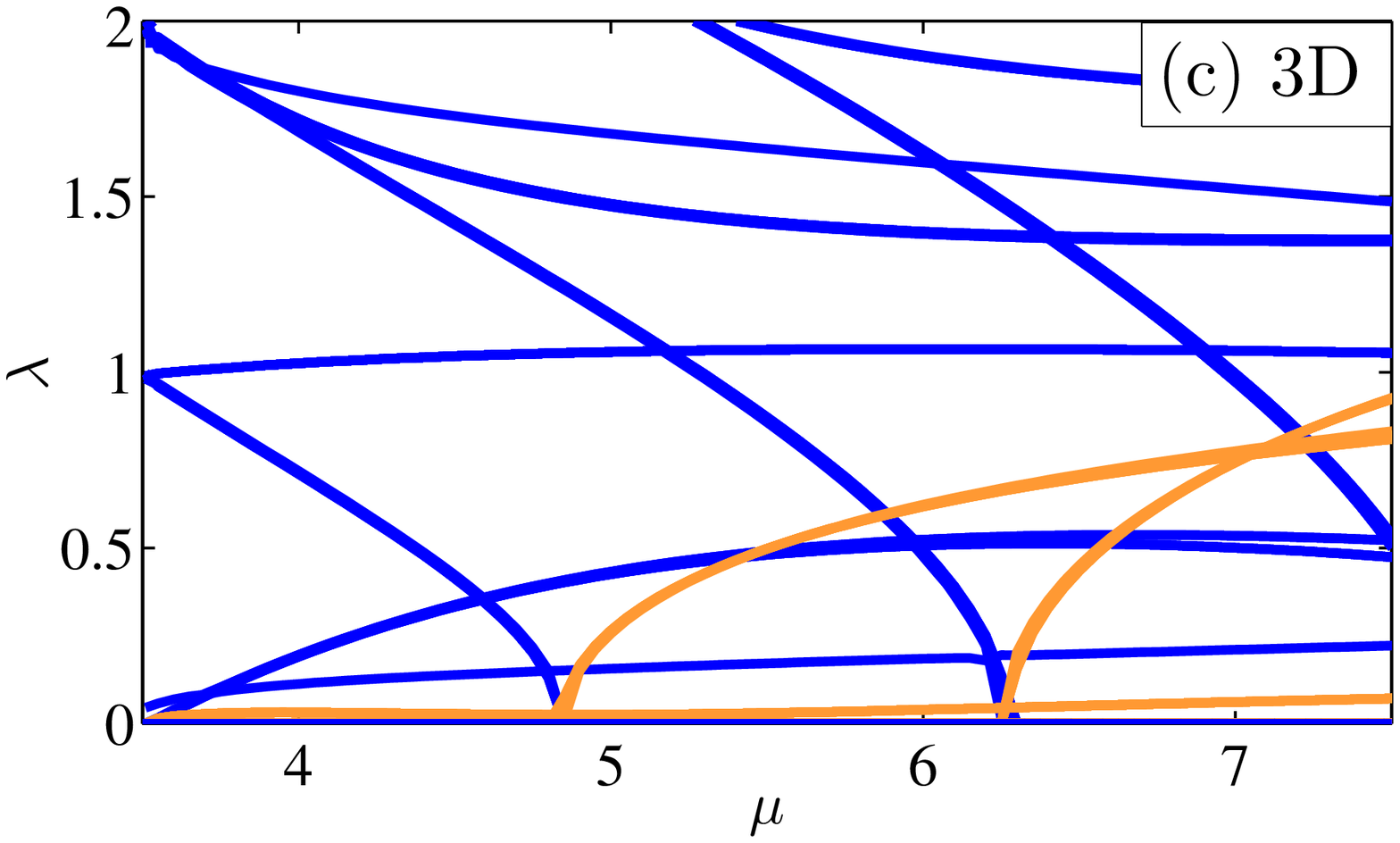}
\caption{(Color online)
Stability spectra for the dark soliton shell as a function of the chemical
potential $\mu$.
(a)-(b) Spectra computed using our effectively 1D method (in the radial
variable, using a spherical harmonic decomposition in the transverse
directions).
Identical results are obtained when using the 2D variant of the method
in cylindrical coordinates (and decomposing only the azimuthal direction
in suitable Fourier modes).
Both 1D and 2D numerics use a spacing of $h=0.01$.
(c) Spectra computed using the full 3D numerics with
spacing $h=0.2$.
%
Orange (grey) lines depict the real part of $\lambda$
(i.e., unstable parts of the eigenvalues),
while blue (dark) lines the imaginary part of $\lambda$.
}
\label{f:spectrum}
\end{center}
\end{figure}

\section{Numerical vs.~analytical results}
\label{results}

%
Let us start by providing
some of the basic
features of the DSS. A typical
DSS state at $\mu=12$ is shown in Fig.~\ref{state}, where its spherical
symmetry is apparent. A radial plot of the wavefunction in the TF limit is depicted in the top panel of Fig.~\ref{LC}. As
the chemical potential is increased the number of atoms (``mass'')
\begin{eqnarray}
N=\int |\psi|^2 dx dy dz= 4 \pi \int_0^{\infty} r^2 |\psi|^2 dr,
\label{mass}
\end{eqnarray}
increases as depicted in the bottom panel of Fig.~\ref{LC}.

%
We now examine the stability of the DSS. The corresponding
spectra illustrating the imaginary part of the eigenvalues
[dark (blue) lines) and the real part [light (orange) lines] are shown
in Fig.~\ref{f:spectrum}. Panel (a) depicts the spectrum obtained via
our numerical method
over a wide range of chemical potentials.
Identical results were obtained when using the method in 2D, based
on a cylindrical coordinate decomposition and a representation of
the azimuthal variable dependence in the corresponding Fourier modes.
For reasons of completeness, a summary of this variant of the method
is presented in the appendix.
Panels (b) and (c) depict
a zoomed in region to contrast the results between these lower
dimensional
 methods and the full 3D numerics. Panel (b) corresponds to the spectrum
as computed by means of the 1D spherical coordinate decomposition
method using spherical harmonics with spatial spacing of $h=0.01$,
while panel (c) depicts the results from direct full 3D numerics
using an spatial spacing of $h=0.2$.
%
%
A close comparison between the panels suggests small discrepancies
of the imaginary parts at Im$(\lambda)=1$ and Im$(\lambda) \approx 0.2$,
and a spurious unstable mode with Re$(\lambda)\approx 0.08$.
These discrepancies (and their amendment as the mesh resolution $h$ decreases ---see below)
are likely due to the large spatial spacing $h$ ---that needs to
be chosen such that the 3D calculations are still ``manageable''---
in which case the DSS radial profile is under-resolved.
It is crucial to note that the 3D calculations are far more
expensive than their lower dimensional (1D or 2D) counterparts
considered above.
%

\begin{figure}[tb]
\begin{center}
\includegraphics[width=7cm]{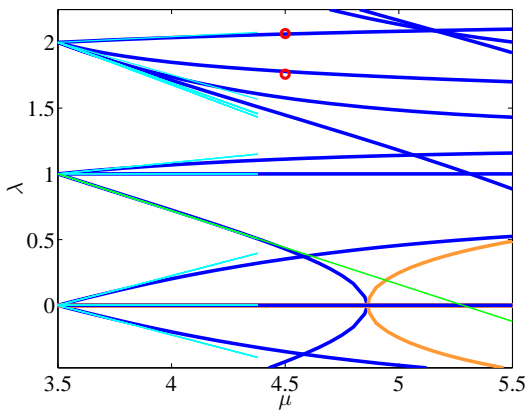}
\caption{(Color online)
Comparison near the linear limit
of the spectrum of the degenerate perturbation method and the
numerical result of the effectively 1D method involving
the spherical harmonic decomposition.
Thick blue (dark) and orange (light) lines depict the imaginary and
real parts of the numerical eigenvalues while the thinner
cyan and green lines depict the theoretical results of the degenerate
perturbation method. The eigenvalue depicted in green is the
one responsible for the destabilization of the DSS.
The large red dots correspond to the two frequencies extracted
from the oscillations of the perturbed DSS depicted in Fig.~\ref{RK4s}.
The larger frequency corresponds to the oscillatory mode of the DSS
while the smaller one corresponds to the breathing mode
of the background density.
}
\label{DPM}
\end{center}
\end{figure}

To confirm that the discrepancies are indeed caused by the mesh resolution, when
we further decrease the lattice spacing to $h=0.15$ (results not shown here) and
recalculate the spectrum in 3D,
the branches at Im$(\lambda)=1$ indeed split into two non-degenerate curves
and the spurious modes at Im$(\lambda) \approx 0.2$ and Re$(\lambda) \approx 0.08$
begin to get suppressed.
This suggests that the discrepancies are indeed attributable to resolution
effects and that the spectrum computed with the effectively
1D numerical method
is indeed accurate.
%

\begin{figure}[tbh]
\begin{center}
\includegraphics[width=6.8cm]{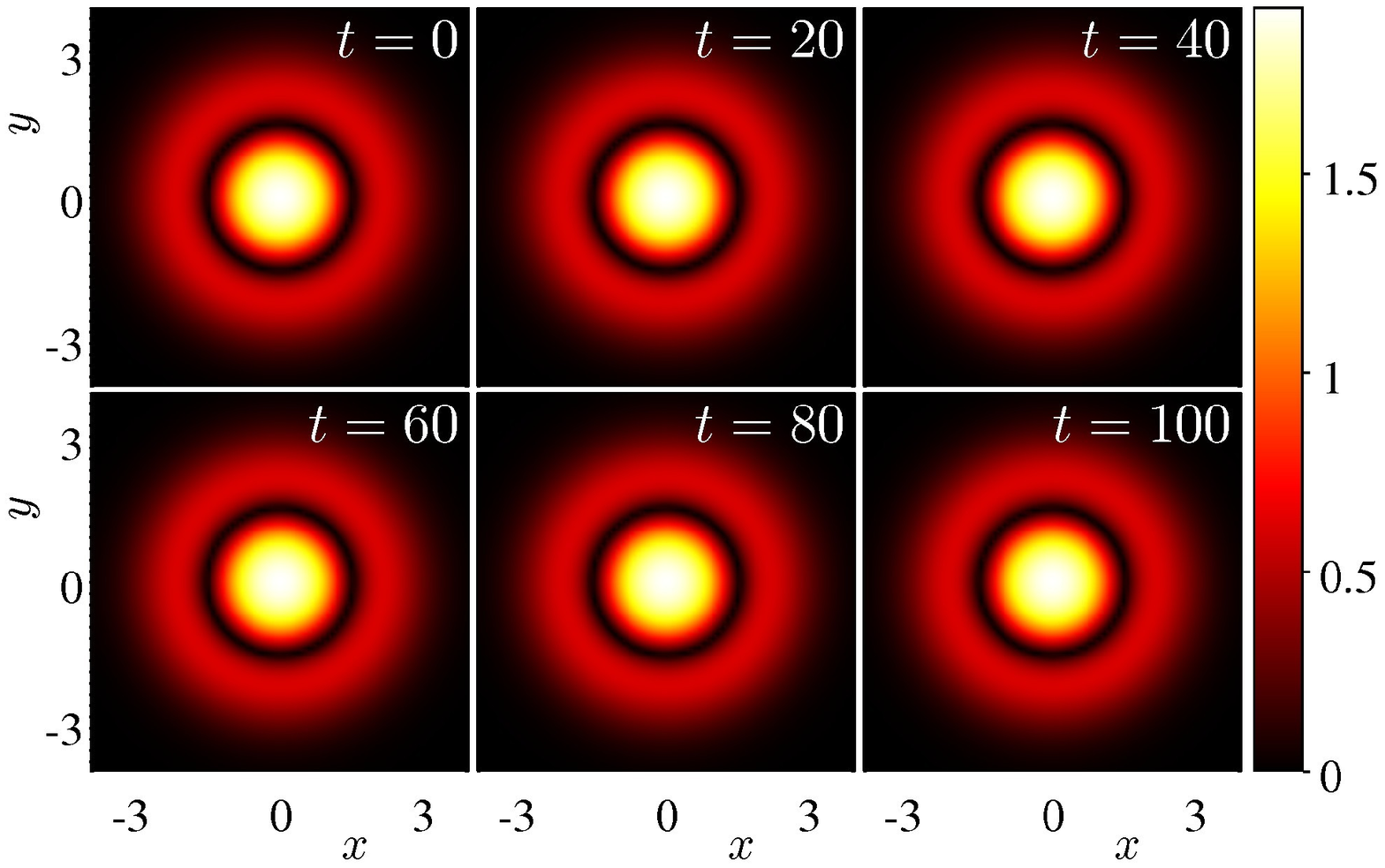}
\includegraphics[width=6.8cm]{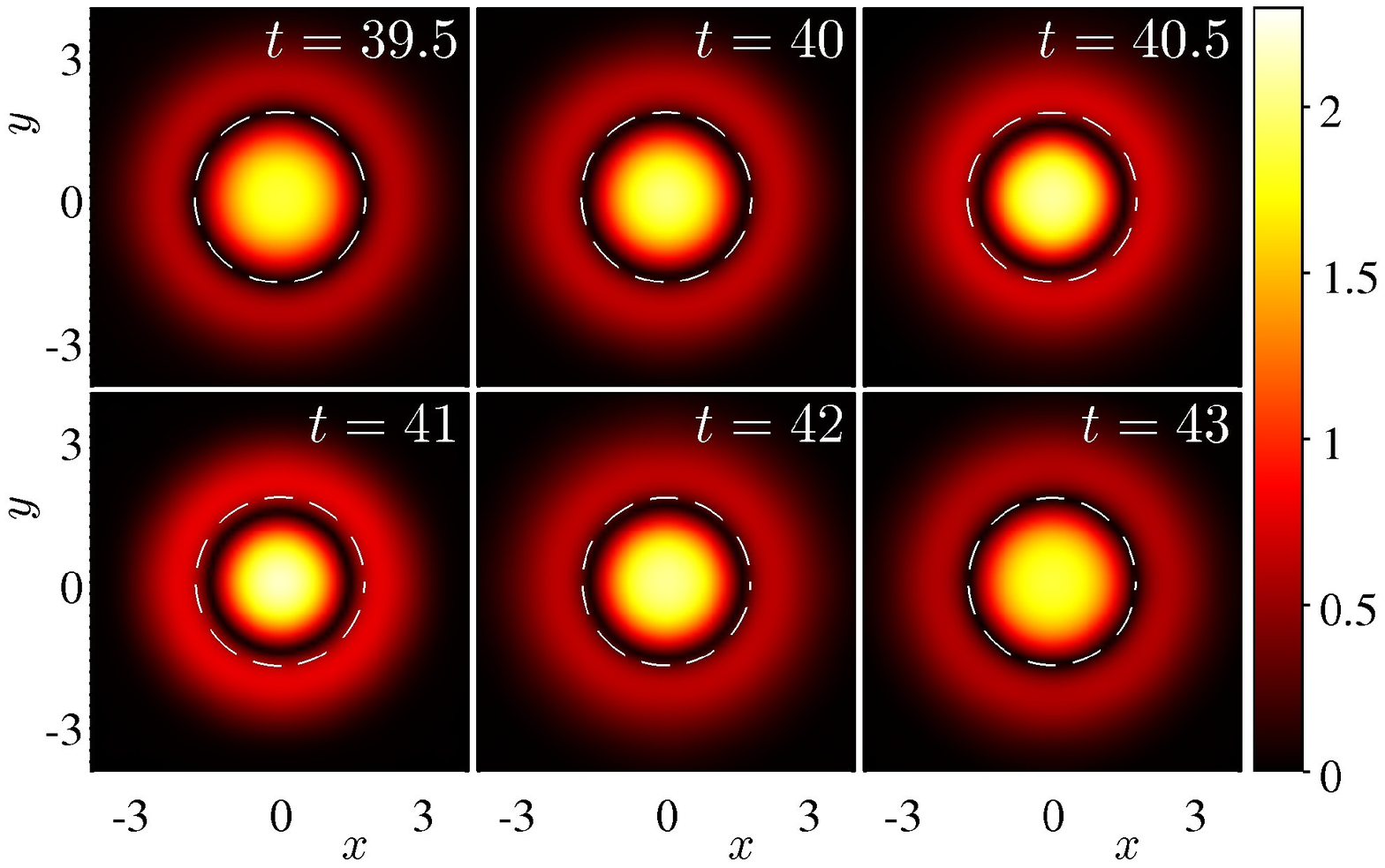}
\includegraphics[width=6.8cm]{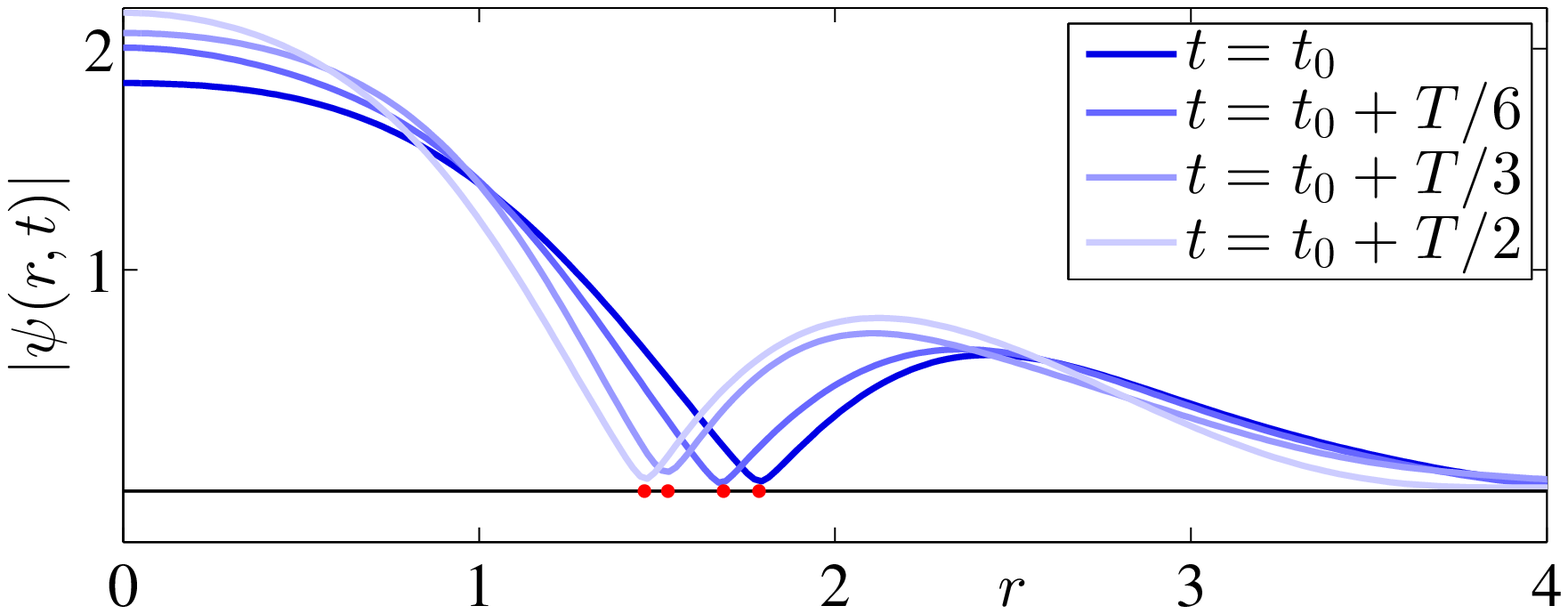}
\includegraphics[width=6.8cm]{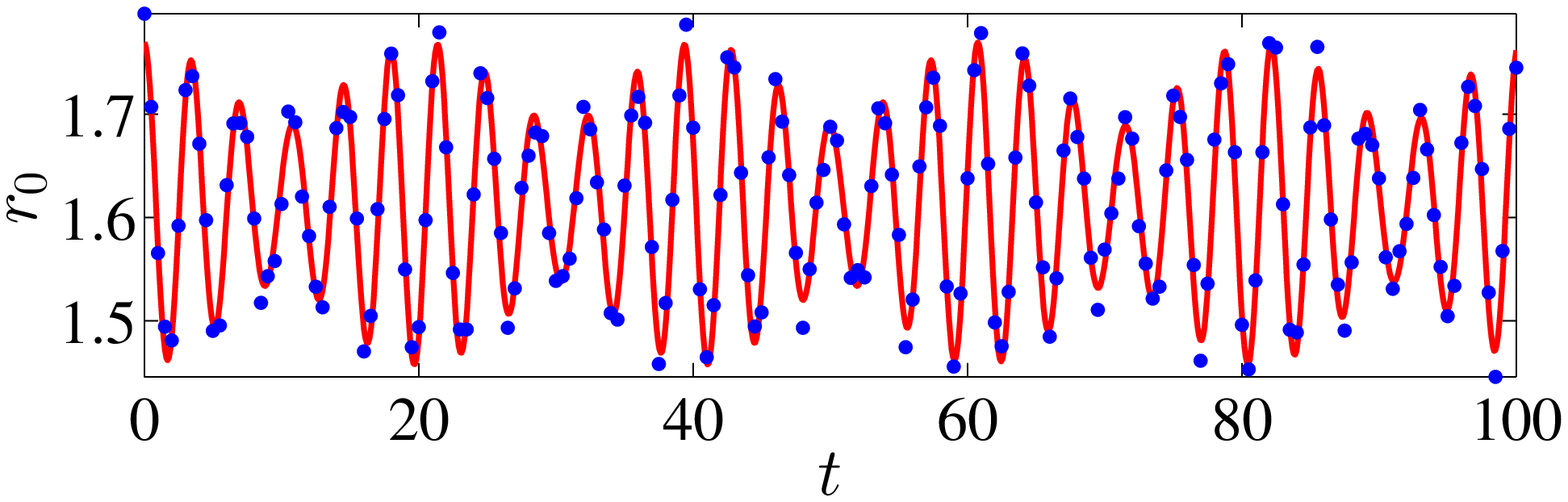}
\caption{
(Color online)
Top set of panels: 2D density cuts ($z=0$)
from the full 3D numerical solution
for $\mu=4.5$. The initial condition corresponds to the stationary
DSS at its equilibrium radius.
%
Second set of panels: Evolution of a DSS shifted from its equilibrium position
undergoing oscillatory motion over one oscillation period $T$.
The white dashed circle denotes the starting location of
the DSS at the beginning of the period.
Third panel: Radial profiles for the oscillating DSS
over half a period.
Bottom panel: Oscillations of the DSS radius vs.~time. Blue dots
represent the radius extracted from full 3D numerics
while the solid red line
is the best (least-square) fit to a linear combination of
oscillations with frequencies $w_1\approx 1.7581$ and $w_2=2.0671$
(see red dots in Fig.~\ref{DPM}) corresponding, respectively, to
the frequencies of
the DSS oscillations and
the breathing mode of the background cloud.
For a movie depicting the stable oscillations of the DSS please
see the supplemental material.
}
\label{RK4s}
\end{center}
\end{figure}

The comparison between the stability spectrum predicted by the
degenerate perturbation method (DPM) and the one obtained
numerically near the linear limit is shown in Fig.~\ref{DPM}.
As the figure shows, the DPM captures
the behavior of the spectrum close to the linear limit.
Let us now confirm the stability of the DSS in this limit.
We have performed a direct numerical integration of the full 3D dynamics
of the DSS for $\mu=4.5$, where it is predicted to be stable
---cf.~Figs.~\ref{f:spectrum} and \ref{DPM}.
This evolution is depicted in the top set of panels of Fig.~\ref{RK4s}.
Also, in this figure, we depict the stable oscillations of a
perturbed DSS by displacing its initial location away from the
equilibrium radius. The evolution for such a DSS performing
oscillations is displayed in the second set of panels in Fig.~\ref{RK4s}.
A movie for this evolution is provided in the supplemental material.
To more clearly visualize these oscillations, we depict on the third
panel of Fig.~\ref{RK4s} the transverse density cuts at four different
stages over half a period of the oscillation.
These density cuts also reveal the excitation of the background
cloud which performs breathing oscillations.
From the density cuts we extracted the (radial) location of the DSS
along the evolution which displays a beating-type quasiperiodic oscillation
as depicted by the blue dots in the bottom panel of Fig.~\ref{RK4s}.
In fact, by (least-square) fitting a linear combination of two harmonic
oscillations (see red line) we extract the frequencies
$w_1\approx 1.7581$ and $w_2=2.0671$. Upon closer inspection, these
frequencies correspond, respectively, to the frequency
of the DSS oscillations and
the frequency of the breathing mode of the background cloud.
These frequencies are also depicted in Fig.~\ref{DPM}
clearly showing that they indeed belong to the stability spectrum
of the stationary DSS state.
It is important to stress that the observed DSS oscillations are
supported by the stability of the DSS at its equilibrium
position. Therefore, despite strong perturbations of the background
that undergoes breathing oscillations, the DSS robustly persists
and performs
stable oscillations about its equilibrium position.

\begin{figure}[tb]
\begin{center}
\includegraphics[width=4.2cm]{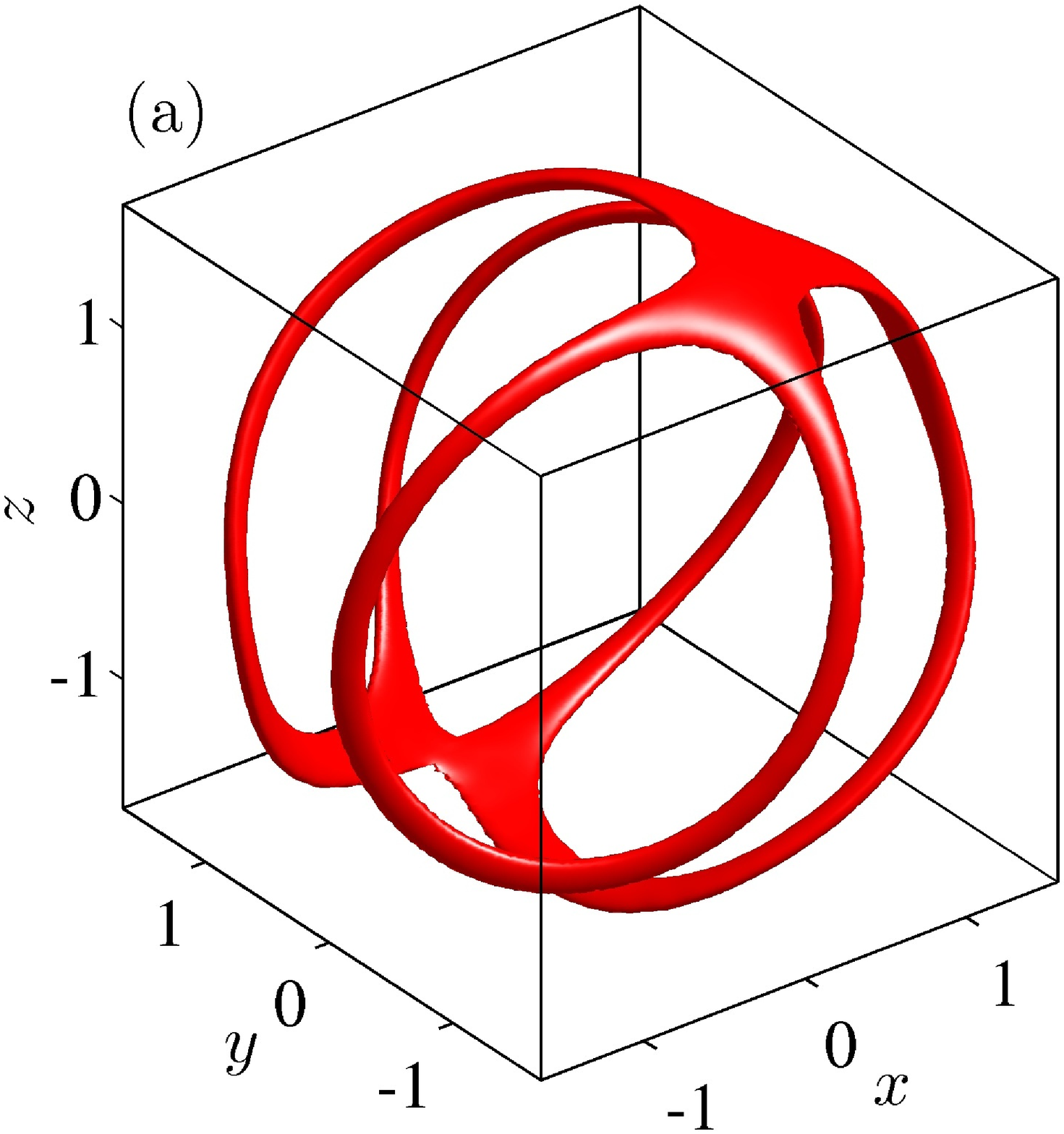}
\includegraphics[width=4.2cm]{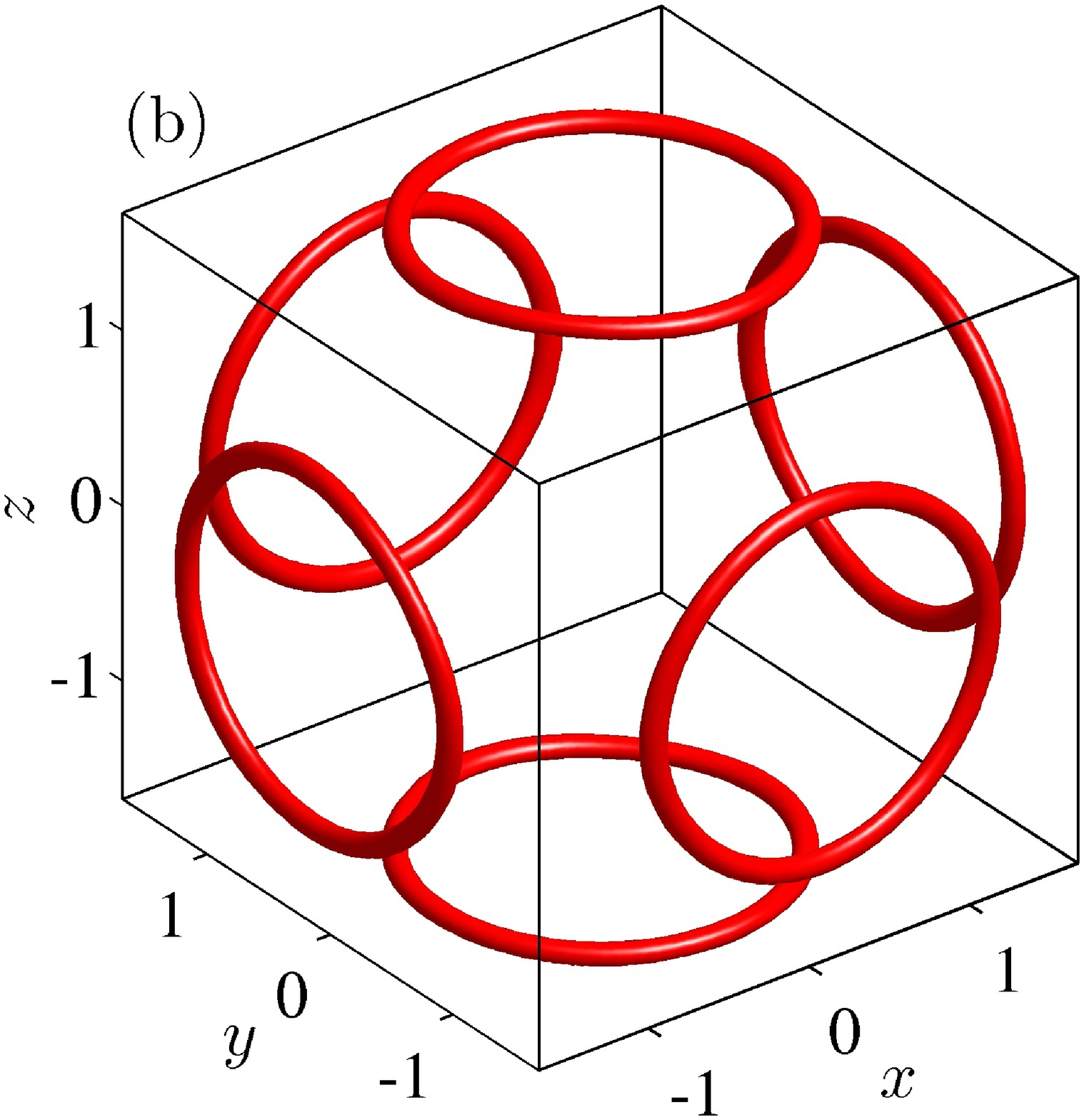}
\caption{(Color online)
The VL6 cage (a) and the VR6 cage (b) that emerge as a result
of the instability of the DSS state.
The states are captured from the
numerical integration of the GPE (\ref{GPE}) dynamics at $\mu=5.8$ and $8.2$,
respectively. In particular, the VR6 cage seems to be robust and appears in
the dynamics for
a wide range of chemical potentials from the onset of the instability
around $\mu=6.2$.
}
\label{VL6VR6}
\end{center}
\end{figure}

As the spectrum indicates, see Figs.~\ref{f:spectrum} and \ref{DPM}, for larger
values of the chemical potential the DSS becomes unstable.
To identify this instability, we have looked at the eigenvectors of the DPM.
There are high degeneracies in the eigenvalues in this case, rendering
harder the identification of
the nature of the instability. Nevertheless, the DPM is
still helpful in
that regard. 
For instance, we know from the eigenvectors
that the mode with the largest slope decreasing at Im$(\lambda)=1$, which causes instabilities,
is a linear combination of ``up'' states, and the ``down'' states are not involved.
Therefore, the instability is due to
a bifurcation rather than a collision
with ``negative'' energy modes; the latter may lead to oscillatory
instabilities, as discussed, e.g., in Ref.~\cite{ourVRfromLinear}.
The former may be associated with symmetry breaking features
and exhibits the instability via real eigenvalue pairs.
Direct numerical integration of Eq.~(\ref{GPE}) for chemical potentials
$\mu=5.8, 8.2$ and 10.6, 
shows that the first instability is
a bifurcation of a six vortex lines (VL6) cage, while the rest of the instabilities
are dominated by a cubic
six vortex rings (VR6) cage. The first instability suggests 
a bifurcation of
VL6 from the interactions of the DSS and a dark soliton state with three plane nodes
perpendicular to the $(x,y)$-plane, 
separated by angles of $\pi/3$; the dark soliton state
(DS3) can be written as
\begin{eqnarray}
|\psi_{\rm{DSm}}\rangle_{\rm{linear}}&\propto& \rho^m \cos(m\phi) e^{-\omega r^2/2},
\end{eqnarray}
with $m=3$ and a linear eigenenergy $(m+\frac{3}{2})\omega$. A two-mode
analysis~\cite{theochar,ourVRfromLinear} using this state and the DSS state yields a prediction of
the bifurcation as occurring at $\mu_c=4.78$, which is in good agreement
with the numerical result
$\mu_c=4.85\pm 0.01$ of Fig.~\ref{f:spectrum}. A similar treatment of the VR6 cage
turns out to be more challenging due to the difficulty in
designing the corresponding
dark soliton state. Instead, we have looked at the DS4 state, which can form an
eight vortex line (VL8) cage. The two-mode analysis then predicts a bifurcation
at $\mu_c=5.97$,
which is in fair agreement with the second instability at
$\mu_c=6.25\pm 0.01$ from Fig.~\ref{f:spectrum}. 
The numerical results, however, show that the dynamically resulting
state is a VR6 cage, not a VL8 cage. Presumably the VL8 cage can also cause
instabilities of the DSS but is less robust than the VR6 cage, possibly due
to the intersections of the vortex lines, which are absent in VR6.
Further analysis with higher values of $m$ (not shown here) also suggests
the progressively lesser relevance of the DSm state in causing instabilities.
On the other hand, the VR6 state seems to be very robust and dominates the
instability in a wide range of chemical potentials from the onset of this
instability around $\mu=6.2$. The VL6 and VR6 cages are shown in Fig.~\ref{VL6VR6}.

\begin{figure}[tb]
\begin{center}
\includegraphics[width=8cm]{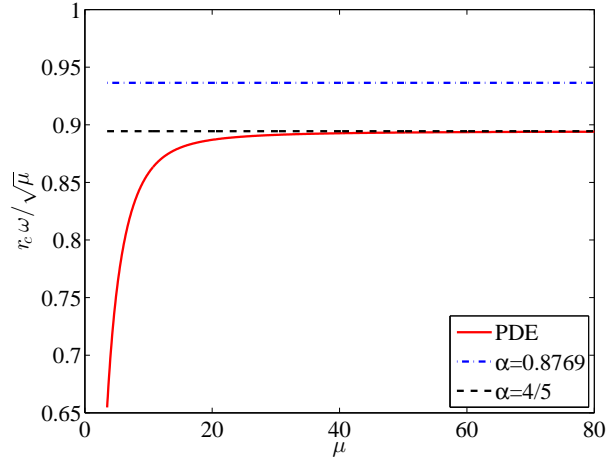}
\caption{(Color online)
The equilibrium radius of the spherical shell dark soliton,
scaled by $\sqrt{\mu}/\omega$, as a function of $\mu$ (solid red line).
Note that the numerical values reach the asymptotic
value of $\sqrt{4/5}$ (dashed black line), as per the second particle picture,
when $\mu$ is large. The first particle
picture 
slightly overestimates $r_c$ (dash-dotted blue line).
%
}
\label{rc}
\end{center}
\end{figure}

Finally, we present the numerical results for the equilibrium location $r_c$
of the DSS as a function of the chemical potential $\mu$ and compare the
results with those of the two particle pictures in the large density limit
(see Sec.~\ref{sec:sub:part}).
A plot of $r_c\omega/\sqrt{\mu}$ as a function of~$\mu$ is depicted in Fig.~\ref{rc}.
Note that the first particle picture predicts a slightly larger equilibrium
$r_c$, while the particle picture based on energy conservation of the
DSS as an adiabatic invariant agrees very
well with the numerical results.
It is interesting to note that the asymptotic
behavior only sets in at about $\mu=30$, deep inside the TF regime.
Although 
this would be a rather computationally demanding parametric range when using 3D (or even 2D)
methods, our effective 1D approach based on a spherical
harmonic decomposition of the angular dependencies, allows us to
reach such large
values of $\mu$.

\section{Conclusions and Future Challenges}
\label{conclusion}

In this work, we have revisited the theme of
spherical shell dark solitons. We have employed analytical methods
to study these structures
both in the vicinity of the linear limit, using a bifurcation
analysis, and in the
Thomas-Fermi limit, where they can be treated as effective particles.
For the numerical simulations, we have used continuation methods,
as well as 
a quasi-one-dimensional
radial computation method, decomposing the angular dependence of
the perturbations in spherical harmonics, to determine their stability.
We have found that spherical shell dark solitons can, in fact, be dynamically
robust, spectrally stable solutions of the 3D Gross-Pitaevskii equation
within a certain interval of chemical potentials (and atom numbers);
this occurs sufficiently close to the linear limit of relatively small chemical potentials.
Their instabilities, emerging when the chemical potential
increases, have been elucidated and their dynamical outcome, 
namely the breakup into vortex ring and vortex line cages, has been
showcased. We believe that these results,
may not only be used to partially explain the
transient observation of these structures in earlier experiments~\cite{harvard} and numerical
simulations~\cite{kominbrand}, but may also
pave the way for their consideration
in future experiments.


Our work suggests a number of interesting future research directions,
concerning, in particular, related coherent structures, 
their physical relevance and
properties. 
For instance, this study suggests large scale stability computations for a wide range of
pertinent steady states with radial symmetry, including ones with different numbers of
radial nodes as well as a number of different external potentials.
%
Furthermore, it would be relevant to generalize and apply the
numerical (and analytical) techniques used herein to multicomponent BEC settings.
For example, one can study 
``symbiotic states'' composed of spherical shell dark-bright solitons,
in the spirit of the dark-bright 2D rings of Ref.~\cite{stockhofe}.
On the other hand, states emerging robustly from the instabilities
of these spherically (or cylindrically) symmetric ones, such as
the vortex ring (and line) cages are worthwhile to investigate further
in their own right.
Efforts along these directions are currently in progress and
will be presented in future publications.

\begin{acknowledgments}

W.W.~acknowledges support from NSF
(Grant Nos.~DMR-1151387 and DMR-1208046).
P.G.K.~gratefully acknowledges the support of
NSF-DMS-1312856, as well as from
the US-AFOSR under grant FA950-12-1-0332,
and the ERC under FP7, Marie
Curie Actions, People, International Research Staff
Exchange Scheme (IRSES-605096).
R.C.G.~gratefully acknowledges the support of NSF-DMS-1309035.
We thank the Texas A\&M University for access to their Ada cluster.

\end{acknowledgments}

\section*{APPENDIX: 2D DECOMPOSITION USING CYLINDRICAL COORDINATES}
The greatest advantage of the 1D method involving the
decomposition into spherical harmonics is that one can
compute the spectrum for large chemical potentials without excessive
computational costs. The reason that typical computations for
large chemical potentials become computationally prohibitive, is that
in this limit the density is large and, thus, the width of the relevant
localized structures (proportional to $1/\sqrt{\mu}$)
becomes much smaller than the domain size (proportional to $\sqrt{\mu}$);
hence, a large number of mesh points is necessary to resolve properly these
configurations.
Nonetheless, it is worth mentioning that, for the projection method in 1D,
the form of states that one can study is also  restricted by
the spherical symmetry; therefore, it is also relevant to employ 
an analog of this method in 2D, so as
to encompass a wider class of states. An example concerns
states with a topological charge $S$ along
the $z$ axis, i.e., stationary states of the form $\psi=\psi_0\, e^{iS\phi}$.
Note that in this framework, one can study states including ---but not limited
to--- the ground-state, planar, ring, or spherical shell dark soliton states,
vortex lines, as well as coplanar or parallel vortex rings
and hopfions~\cite{hopfion},
in both isotropic and anisotropic traps.

In 2D, and for 
cylindrical coordinates $(\rho,\phi,z)$, 
the Laplacian is decomposed as follows:
\begin{equation}
\nabla^2 f =\Delta_H f+\frac{\Delta_G f}{\rho^2},
\end{equation}
where its components read:
%
\begin{eqnarray}
\Delta_H f &=&\frac{1}{\rho} \frac{\partial}{\partial \rho} \left(\rho \frac{\partial f}{\partial \rho}\right)+\frac{\partial^2 f}{\partial z^2},\\
\Delta_G f &=&\frac{\partial^2 f}{\partial \phi^2}.
\end{eqnarray}
%
%
The $\Delta_G$ operator has eigenstates $\{e^{im\phi}\}$ with
eigenvalues $-m^2$, i.e.,
\begin{equation}
\Delta_G e^{im\phi} =-m^2e^{im\phi}.
\end{equation}
%
A stationary state $\psi_0(\rho,z)$, defined in the domain
$\rho \in [0,\infty) \times z \in \mathbb{R}$, and bearing a
topological charge $S$, 
satisfies the following equation:
\begin{equation}
-\frac{1}{2} \Delta_H \psi_0+\frac{S^2}{2\rho^2}\psi_0
+V(\rho,z)\psi_0 +| \psi_0 |^2 \psi_0-\mu \psi_0=0,
\end{equation}
%
%

As 
in the 1D case, 
we construct the
linear stability 
problem as follows.
Let $\psi_0$ 
a rotationally symmetric stationary state, up to a topological charge $S$, along the $z$ axis
perturbed using the complete basis of $\{e^{im\phi}\}$: 
\begin{equation}
\psi
=e^{iS\phi} \left[ \psi_0+\sum\limits_{m} \left[ a_{m}(\rho,z,t)e^{im\phi}+b_{m}^*(\rho,z,t)e^{-im\phi} \right] \right].
\nonumber
\end{equation}
%
Substituting this expansion into
the GPE of Eq.~(\ref{GPE}), and upon linearizing
and matching the basis expansion
on both sides, one can see that the different $m$ modes are mutually
independent. An equivalent derivation as in 1D shows that $\lambda$ are
the eigenvalues of the matrix
$$M=
\begin{pmatrix}
M_{11} & M_{12} \\
M_{21} & M_{22}
\end{pmatrix},
$$
where
$$
\begin{array}{rcl}
M_{11}&=&-i\left(-\frac{1}{2} \Delta_H + \frac{(m+S)^2}{2\rho^2}  +V  +2|\psi_0|^2-\mu\right),\\[2.0ex]
M_{12}&=&-i\psi_0^2,\\[1.0ex]
M_{21}&=&i\psi_0^{*2},\\[1.0ex]
M_{22}&=&i\left(-\frac{1}{2} \Delta_H + \frac{(m-S)^2}{2\rho^2}  +V  +2|\psi_0|^2-\mu\right).
\end{array}
$$
As before, one can therefore compute the full spectrum by computing each
mode $m$ independently and then putting them together. In the results
based on this method and presented in Sec.~\ref{results}, we
use $m=0,1,2,...,5$.
%
It is worth commenting here that the matrix for the 2D eigenvalue problem
with a charge $S$ is less symmetric
than matrices
corresponding to the 1D projection and the 2D case with no charge.
Therefore, the full 2D problem with charge appears to demand more computational work.
However, it should be noted that, in the 2D case, the eigenvalues of the full matrix $M$
corresponding to the projections over the $+m$ and $-m$ modes are related via an orthogonal
transformation and, thus, the two set of eigenvalues are complex conjugates
of each other. Therefore, it is only necessary to compute the non-negative $m$
set and, 
hence, the charged 
and the uncharged states actually have similar computational complexity.


\end{document}